\documentclass[twocolumn,showpacs,prb,floatfix]{revtex4}

\usepackage{graphicx}
\usepackage{bm}
\begin{document}

\title{Molecular dynamics with quantum heat baths: results for
nanoribbons and nanotubes}

\author{Jian-Sheng Wang}%
\author{Xiaoxi Ni}%
\author{Jin-Wu Jiang}%
\affiliation{Department of Physics and Center for Computational Science and Engineering, National University of Singapore, Singapore 117542, Republic of Singapore}

\date{19 July 2009}

\begin{abstract}
A generalized Langevin equation with quantum baths (QMD) for thermal
transport is derived with the help of nonequilibrium Green's function
(NEGF) formulation. The exact relationship of the quasi-classical
approximation to NEGF is demonstrated using Feynman diagrams of the
nonlinear self energies.  To leading order, the retarded self energies
agree, but QMD and NEGF differ in lesser/greater self energies.  An
implementation for general systems using Cholesky decomposition of the
correlated noises is discussed.  Some means of stabilizing the
dynamics are given.  Thermal conductance results for graphene strips
under strain and temperature dependence of carbon nanotubes are
presented.  The ``quantum correction'' method is critically examined.
\end{abstract}

\maketitle

\section{Introduction}

Molecular dynamics\cite{MDbook} (MD) has been used as one of the most
important simulation tools to study a variety of problems from
structure to dynamics.  In particular, MD is routinely used for
thermal transport problems\cite{lepri03,mcgaughey06}.  MD is versatile
and can handle with ease any form of classical interaction forces.
The computer implementation of the algorithms is straightforward in
most cases.  However, there is one essential drawback in MD -- it is
purely classical, thus it is unable to predict quantum behavior.  Of
course, in situations like high temperatures and atoms other than
hydrogen, MD gives good approximation.  This is not the case for very
small systems like nanostructures at low temperatures.  For lattice
vibrations, the relevant temperature scale is the Debye temperature,
which is quite high for carbon-based materials.  Thus, even the room
temperature of 300 K is already considered a low temperature.  One
uses MD anyway in cases where quantum effects might be important, for
lack of better alternative approaches.

Recently, it was proposed that MD can be augmented with a quantum heat
bath to at least partially take into account the quantum effect
\cite{wangPRL07}. We'll refer to this new molecular dynamics as
QMD. The proposed generalized Langevin dynamics with correlated noise
obtained according to Bose distribution has the important features
that it gives correct results in two special limits, the 
low-temperature ballistic limit and high-temperature diffusive limit.  It
is one of the very few methods that ballistic to diffusive transport
can be studied in a single unified framework.  The QMD should be most
accurate for systems with strong center-lead couplings.  The
non-Markovian heat baths have an additional advantage in comparison
with the usual Langevin or Nos\'e-Hoover heat baths in that the baths
(the leads) and the systems can be connected seamlessly without
thermal boundary resistance.

In this paper, we follow up the work of ref.~\onlinecite{wangPRL07} to
give further details and calculations on large systems.  We first
discuss the implementation of the colored noises with several degrees
of freedom.  We then discuss the problem of instabilities we
encountered in carrying out the simulation.  Various ways of
overcoming this difficulty are suggested and tested.  We analyze the
proposed dynamics and compare it with the exact nonequilibrium Green's
function (NEGF) method\cite{wangrev08} in terms of Feynman diagrams and
nonlinear self-energies.  Here we show that in the ballistic case, the
generalized Langevin dynamics and NEGF are completely equivalent.
Using this analysis, it is also clear that at high temperatures NEGF
and Langevin dynamics should agree for nonlinear systems.  We then
report some of the simulation results on nanoribbons (with periodic
boundary condition in the transverse direction) and nanotubes.  We
comment on one popular method of ``quantum correction'' and point out
its shortcomings and inconsistencies.  We conclude in the last
section.

\section{Generalized Langevin dynamics and implementation details}
The derivation of the generalized Langevin equation for junction
systems was given in ref.~\onlinecite{wangPRL07}, see also
refs.~\onlinecite{adelman76} and \onlinecite{dharJSP06}.  Here we
present a much faster derivation using the results of NEGF, following
the notations introduced in ref.~\onlinecite{wangrev08}.  The starting
point is the set of quantum Heisenberg equations of motion for the
leads and center,
\begin{eqnarray}
\label{eqcenter}
\ddot{u}^C &=& F^C -V^{CL}u^L - V^{CR}u^R, \\
\ddot{u}^\alpha &=& -K^\alpha u^\alpha -V^{\alpha C}u^C,\qquad \alpha=L,R.
\end{eqnarray}
$u^\alpha$ is a vector of displacements in region $\alpha$ away from
the equilibrium positions, multiplied by the square root of mass of
the atoms.  The leads and the coupling between the leads and center
are linear; while the force in the center, $F^C(u^C) = -K^Cu^C + F_n$,
is arbitrary.  We eliminate the lead variables by solving the second
equation and substituting it back into the first equation.  The
general solution for the left lead is
\begin{equation}
\label{equL}
u^L(t) = u_0^L(t) + \int^t \!g^r_L(t-t')\, V^{LC} u^C(t')\, dt',
\end{equation}
where $g^r_L(t)$ is the retarded Green's function of a free left lead
with the spring constant $K^L$, satisfying $\ddot{g}^r_L(t) + K^L
g^r_L(t) = -\delta(t) I$, $g^r_L(t) = 0$ for $t\leq 0$.  Its Fourier
transform is given by $[(\omega + i \eta)^2 -K^L]^{-1}$, where $\eta
\to 0^{+}$ is an infinitesimal positive quantity. $u_0^L(t)$ satisfies
the homogeneous equation of the free left lead:
\begin{equation}
\ddot{u}_0^L + K^L u^L_0 = 0.
\end{equation}
$g^r_L(t)$ and $u_0^L(t)$ are associated with the ``free'' lead in the
sense that leads and center are decoupled, as if $V^{CL}=0$.  This is
consistent with an adiabatic switch-on of the lead-center
couplings. The right lead equations are similar.  Substituting
Eq.(\ref{equL}) into Eq.~(\ref{eqcenter}), we obtain
\begin{equation}
\label{eq-langevin}
\ddot{u}^C = F^C  - \int^t \Sigma^r(t-t')u^C(t') dt' + \xi,
\end{equation}
where $\Sigma^r = \Sigma_L^r + \Sigma_R^r$, $\Sigma_\alpha^r =
V^{C\alpha}g_\alpha^r(t)V^{\alpha C}$, is the self-energy of the
leads, and the noise is defined by $\xi_\alpha(t) =
-V^{C\alpha}u^\alpha_0(t)$, $\xi = \xi_L + \xi_R$.

The most important characterization of the system is the properties of
the noises.  This is fixed by assuming that the leads are in
respective thermal equilibrium at temperature $T_L$ and $T_R$. It is
obvious that for a set of coupled harmonic oscillators, there is no
thermal expansion effect, $\langle u_0^\alpha(t) \rangle= 0$, thus
$\langle \xi_\alpha \rangle = 0$.  The correlation function of the
noise is
\begin{eqnarray}
\langle \xi_L(t) \xi_L^T(t') \rangle &= &
V^{CL} \langle u_0^L(t) u_0^L(t')^T \rangle V^{LC} \nonumber \\
&=& V^{CL} i\hbar g_L^{>}(t-t') V^{LC} \nonumber\\
&=& i\hbar \Sigma_L^{>}(t-t'), \label{eqnoise1}
\end{eqnarray}
where the superscript $T$ stands for matrix transpose.  We have used the
definition of greater Green's function and self-energy of the free
left lead\cite{wangrev08}.  We assume that the noises of the left lead
and right lead are independent.  Since the noises $\xi_\alpha(t)$ are
quantum operators, they do not commute in general.  In fact, the
correlation in the reverse order is given by the lesser self-energy:
\begin{eqnarray}
\label{eqnoise2}
\langle \xi_L(t') \xi_L^T(t) \rangle^T =  i\hbar \Sigma_L^{<}(t-t').
\end{eqnarray}

Equation~(\ref{eq-langevin}) together with the noise correlations
Eq.~(\ref{eqnoise1}) and (\ref{eqnoise2}) is equivalent to NEGF
approach.  For the quantum Langevin equation, it is not sufficient to
completely characterize the solution by just the first and second
moments of the noises.  We need the complete set of $n$-point
correlators $\langle \xi(t_1) \xi(t_2) \cdots \xi(t_n) \rangle$, which
is in principle calculable from the equilibrium properties of the lead
subsystem \cite{ford-kac-mazur65}.  It is very difficult to solve the
dynamics unless the nonlinear force $F_n$ is zero. Thus, for computer
simulation in the quantum molecular dynamics approach, we have
replaced all operators by numbers and a symmetrized noise, $i\hbar
\frac{1}{2}\bigl[\Sigma^{<}(t) + \Sigma^{>}(t)\bigr]=i\hbar
\bar{\Sigma}(t)$, is used.  This is known as quasi-classical
approximation in the literature\cite{schmid82,weiss99}.

\subsection{Implementation}
The formula for the noise spectrum of the left lead is
\begin{equation}
 F[\omega] =  i\hbar \bar{\Sigma}_L[\omega] =
\hbar  \left[ f(\omega) + \frac{1}{2} \right]  \Gamma_L[\omega],
\end{equation}
where $f(\omega) = 1/[e^{\hbar \omega/(k_B T_L)} -1]$ is the Bose
distribution function, and $\Gamma_L[\omega] = i
\bigl(\Sigma^r_L[\omega] - \Sigma^a_L[\omega]\bigr)$.  The right lead
is analogous.  The surface Green's functions $g^r_\alpha$ are obtained
using an iterative method.\cite{lopez-sancho85,wangrev08} To generate
the multivariate gaussian distribution with an arbitrary correlation
matrix, we use the algorithm discussed in ref.~\onlinecite{fishman96}.
That is, we do $Z = c X$, where $X$ is a complex vector following
standard uncorrelated gaussian with unit variance, one for each
discretized frequency, while $c c^T = F[\omega]$, and $c$ is a lower
triangular real matrix.  $c$ is obtained by Cholesky
factorization\cite{golub96} from a Lapack routine {\tt dpotrf( )}.
The Cholesky decomposition is performed only once.  The frequency
array of lower triangular matrices $c$ is stored.  The Fourier
transform of $Z$ gives the noise in time domain, which is obtained
using a fast Fourier transform algorithm. Further details are given in
ref.~\onlinecite{wangPRL07}.

\subsection{Overcoming instability}

We have implemented the second generation reactive empirical bond
order (REBO) Brenner potential\cite{brenner02} for carbon with the
special restriction that the coordination numbers are always three.
This is valid for carbon nanotubes and graphene sheets with small
vibrations in thermal transport.  We found that a naive implementation
of the QMD in higher dimensions unstable.  The atoms close to the
leads have a tendency to run away from the potential minima and go to
infinity.  Several ways were tried to stabilize the system.

(1) Instead of integrating over the coordinates $u^C(t)$ in the memory
kernel, we can perform an integration by part, and consider
integrating over velocity.  This form of the generalized Langevin
equation resembles more of the standard Langevin equation of velocity
damping, but there will be an extra force constant term, as follows:
\begin{equation}
\label{eq-langevin-2}
\ddot{u}^C = F^C  + \lambda \Gamma(0)u^C - \int^t \Gamma(t-t')\dot{u}^C(t') dt' + \xi,
\end{equation}
where $\Gamma(t)$ is defined by Eq.~(\ref{eq-Gamma}) below.  We
introduce a parameter $\lambda$, which should take the value 1, but
using a smaller value can stabilize the system.  However, $\lambda
\neq 1$ introduces boundary resistance.

(2) We scale up the force constants of the leads by a factor of $f$.
This broadens the lead spectra to be closer to white noise, thus
better damping.

(3) We add an additional onsite force on each atom, with a linear
force constant $K_{\rm onsite}$, as well as a small $u^4$ nonlinear
force.  This breaks the translational invariance so that the atoms are
fixed near their equilibrium positions.

(4) We smooth the noise spectrum by choosing a small number of points,
say 100 sampling points in frequency.  We add an artificial damping,
$e^{-\epsilon t}$ to Eq.~(\ref{eq-Gamma}).

(5) We implemented three algorithms: velocity Verlet, fourth order
Runge-Kutta, and an implicit two-stage fourth-order
Runge-Kutta.\cite{IRK94}.

Not all of the measures are effective.  We feel perhaps the most
important point is (4).  The extra parameters $\lambda$, $f$, $K_{\rm
onsite}$, and $\epsilon$ will be stated when discussing the results.

\section{delta-peak singularities in lead self-energy}

\begin{figure}
\includegraphics[width=\columnwidth]{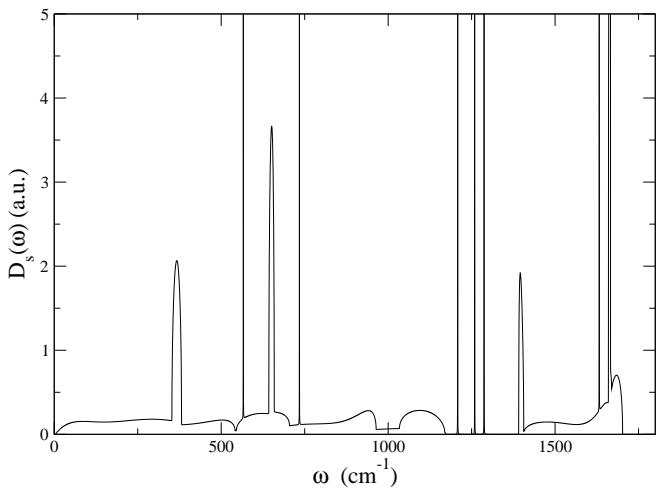}%
\caption{\label{fig-selfepeak}The surface density of states
$D_s(\omega)$ vs.~frequency for a $(n,2)$ zigzag graphene strip with
(1,2) of 8 atoms as a repeating unit cell.  The delta peaks are
located at 566, 734, 1208, 1259, 1287, and 1632 cm$^{-1}$.  The rest
of the peaks do not diverge as $\eta \to 0$.}
\end{figure}

\begin{figure}
\includegraphics[width=\columnwidth]{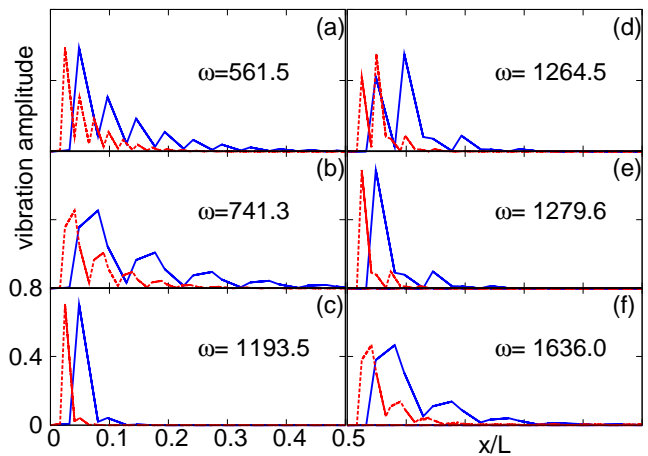}
\caption{(color online). Normalized vibration amplitudes vs.~reduced
coordinate $x$ of each carbon atom. From (a) to (f) are six edge modes in (10,
2) graphene strip (blue solid line) and (20, 2) graphene strip (red
dotted line). The frequency $\omega$ for each mode given in the figure
is in cm$^{-1}$.}
\label{fig_fixbc_edgestate}
\end{figure}

For a one-dimensional (1D) harmonic chain with a uniform spring constant
$K$, the lead self-energy is given by $\Sigma^r[\omega] = -K \lambda$,
where $\lambda$ satisfies $K\lambda + (\omega+i\eta)^2 -2K +
K/\lambda=0$ and $|\lambda|<1$.  Both the real part and imaginary part
are smooth functions of the angular frequency $\omega$.  However, this
is not true in general.  For sufficiently complex leads, we find
$\delta$-function-like peaks on an otherwise smooth background, see
Fig.~\ref{fig-selfepeak}.  What is plotted in Fig.~\ref{fig-selfepeak}
is the surface density of states defined according to
\begin{equation}
D_s(\omega) = - \frac{2\omega}{\pi} {\rm Im} {\rm Tr}\,g^r_L[\omega].
\end{equation}
The above formula gives the bulk phonon density of states if $g^r$ is
replaced by the central part Green's function $G^r$.  The peaks in
Fig.~\ref{fig-selfepeak} are not numerical artifacts, but real
singularities in the semi-infinite lead surface Green's function or
self-energy.  If we omit them, the identity (a special case of the
Kramers-Kronig relation)
\begin{equation}
\Sigma^r[\omega=0] = \int_{-\infty}^{+\infty} \frac{d\omega}{\pi}
\frac{{\rm Im}\, \Sigma^r[\omega]}{\omega}
\end{equation}
will be violated.  These sparks are indeed $\delta$ functions. As the
small quantity $\eta$ in $(\omega + i\eta)^2$ decreases, the peaks
become higher and narrower, but the integral in a fixed interval
around the peak remains constant.  These peaks are not related to the
van Hove singularities of bulk system density of states, as the
locations of the peaks are not at these associated with zero group
velocities.  The singularities of the self-energy can be approximated
as proportional to
\begin{equation}
\frac{1}{\omega - \omega_0 + i\eta} \approx P \frac{1}{\omega-\omega_0}
 -i\pi \delta(\omega -\omega_0),
\end{equation}
where $P$ stands for principle value.

To interpret these peaks, we did a calculation of the vibrational
eigenmodes for a finite system. We find localized modes with
frequencies matching that of the delta peaks in $D_s(\omega)$. We use
``General Utility Lattice Program'' (GULP)\cite{Gale} to calculate the
phonon modes of graphene strip, with fixed boundary condition in $x$
direction and periodic boundary condition in $y$ direction. These
boundary conditions are the same as that in the calculation of lead
surface Green's function. Six localized modes are found in ($n$, 2)
graphene strips. Fig.~\ref{fig_fixbc_edgestate} shows the normalized
vibrational amplitude of each atom in all six localized modes of
(10,2) (blue solid) and (20, 2) graphene strip (red dotted). In these
modes the vibrational amplitude decreases exponentially to zero from
edge into the center.  The frequency of these localized modes are
561.5, 741.3, 1193.5, 1264.5, 1279.6, and 1636.0 cm$^{-1}$. These
values match very well to that of the delta peaks in $D_s(\omega)$
shown in Fig.~\ref{fig-selfepeak}. These frequency values are the same
in (10, 2) and (20, 2) graphene strip.
After a critical distance $L_{c}$ from the edge, the vibration
amplitude decreases to zero. $L_{c}$ is the same in (10, 2) and (20,
2) graphene strip. So these localized modes are relatively more
`localized' in longer graphene strip as shown in
Fig.~\ref{fig_fixbc_edgestate}.  Because of their localizing property,
these modes are important in thermal transport.  There are localized
modes both at the edges of leads and edges of center.  They have
opposite effect on the thermal conductance. (1) Localized modes at the
edges of leads are beneficial for thermal conductance.  Because in
these modes, atoms at the edges of leads have very large vibration
amplitude. As a result, thermal energy can transport from leads into
center more easily. (2) Localized modes at the edges of the center
have negative effect on thermal conductance. In these modes, only
outside atoms have large vibrational amplitudes while inside atoms
have small vibration amplitudes or even do not vibrate at all.  So
thermal energy are also localized at the edges, making it difficult to
be transported from one end to the other end. The effects of (1) and
(2) are such that the net result is equivalent to a perfect periodic
system without boundary resistance.

The localized modes are a consequence of dividing the infinite system
abruptly and artificially into leads and center.  Implementing these
delta-peak singularities in a QMD simulation is impossible, since
these modes are not decaying in time for the real-time self-energy
$\Sigma^r(t)$.  Thus, we are forced to remove these peaks from the
imaginary part of $\Sigma^r$, and reconstruct a real part using the
Hilbert transform from the imaginary part with the delta peaks
removed.  The damping kernel for the QMD dynamics is computed from
(for $t>0$)
\begin{equation}
\label{eq-Gamma}
\Gamma(t) = -\int_t^{+\infty}\!\!\!\!\Sigma^r(t') dt' = \int_{-\infty}^{+\infty} \frac{d\omega}{\pi}
\frac{{\rm Im} \Sigma^r[\omega]}{ - \omega + i\epsilon} e^{-i\omega t - \epsilon t}.
\end{equation}
In practice, the removal of the peaks is done by choosing a small
$\eta$ $(\approx 10^{-8})$.  Since the sampling of $\omega$ is at a
finite spacing, typically with about $10^2$ points, we almost always
miss the peaks if $\eta$ is small.

In calculating the ballistic transmission through Caroli formula, the
omission of the delta peaks at a set of points of measure zero has no
consequence.  However, the existence of the singularities is also
reflected through the real part of the self-energy.  If the real part
uses the Hilbert transformed version with the delta peaks omitted, the
transmission coefficient $T[\omega]$ will not be flat steps as
expected for a perfect periodic system.  Thus, removing the delta
peaks consistently means we are using a lead that is modified from the
original one.

\section{Difference between QMD and NEGF: A Feynman diagrammatic analysis}
In this section, we give an analysis of the difference between fully
quantum-mechanical NEGF and the quasi-classical generalized Langevin
dynamics.  A similar result was presented in
ref.~\onlinecite{luwang09} briefly for the case of electron-phonon interaction.
The starting point is a formal solution of
Eq.~(\ref{eq-langevin}) with the quasi-classical approximation and a
symmetrized correlation matrix for the noise:
\begin{equation}
\label{eq-u-expand}
u(t) = - \int G^r(t,t') \bigl[ \xi(t') + F_n(t') \bigr] dt',
\end{equation}
where we have omitted the superscript $C$ on $u$ for simplicity, $G^r$
is the retarded Green's function of the central region for the
ballistic system (when $F_n=0$).  We have also left out a possible
term satisfying a homogeneous equation [Eq.~(\ref{eq-langevin}) when
$\xi=F_n=0$] and depending on the initial conditions.  Physically,
such term should be damped out.  Provided that the central part is
finite, such term should not be there and this is consistent with the
fact that the final results are independent of the initial distribution of
the central part in steady states.

We consider the expansion of the nonlinear force of the form
\begin{equation}
(F_n)_i = - \sum_{j,k} T_{ijk} u_j u_k - \sum_{j,k,l} T_{ijkl} u_j u_k u_l,
\end{equation}
where $T_{ijk}$ and $T_{ijkl}$ are completely symmetric with respect
to the permutation of the indices.  From repeated substitution of
Eq.~(\ref{eq-u-expand}) back into itself, we can see that $u(t)$ is
expressed as polynomials of $G^r$ and $\xi$.  The correlation
functions of $u$ can then be calculated using the fact that the noise
is gaussian, and Wick's theorem applies.  It is advantageous to define
two types of (quasi-classical) Green's functions, as
\begin{eqnarray}
\label{eq-def1}
-\frac{i}{\hbar} \bigl\langle u(t) u^T(t') \bigr\rangle & \equiv &
\bar  \mathcal{G}_n(t,t') \\
\label{eq-def2}
\frac{i}{\hbar} \bigl\langle u(t) \xi_L^T(t') \bigr\rangle & \equiv &
\int \mathcal{G}_n^r(t,t'') \bar{\Sigma}_L(t'',t') dt''.
\end{eqnarray}

The energy current is calculated by the amount of decrease of energy
in the left (or right) lead:
\begin{equation}
I_L = - \Bigl\langle \frac{H_L}{dt} \Bigr\rangle = - \frac{\partial}{\partial t}
\bigl\langle {u^C(t)}^T V^{CL} u^L(t') \bigr\rangle\Big|_{t=t'=0}.
\end{equation}
We can replace $u^L$ with the solution, Eq.~(\ref{equL}).  Going into
the Fourier space and some algebraic manipulation, we can write
\begin{equation}
\label{eq-cl-IL}
I_L = - \int_{-\infty}^{+\infty} \frac{d\omega}{2\pi}
\hbar \omega\, {\rm Tr} \bigl(
\mathcal{G}^r_n[\omega] \bar{\Sigma}_L[\omega] +
\bar{\mathcal{G}}_n[\omega] \Sigma_L^a[\omega]  \bigr),
\end{equation}
where the Fourier transform is defined in the usual way, e.g.,
$\bar{\mathcal{G}}[\omega] = \int_{-\infty}^{+\infty}
\bar{\mathcal{G}}(t) e^{i\omega t} dt$.  The above equation has the
same form as the NEGF one, provided that we can identify the
quasi-classical Green's functions defined in Eq.~(\ref{eq-def1}) and
(\ref{eq-def2}) with the quantum ones, $\bar{G} = \frac{1}{2}(G^{>} +
G^{<})$ and $G^r$.  [It looks slightly different from the expression
of Eq.~(5) in ref.~\onlinecite{wang-wang-zeng-06}, where only $G^{<}$
appears.  There is an error in that paper.  One should take the real
part of that expression, or add its complex conjugate. By doing this,
we obtain a symmetrized expression with respect to $G^{<}$ and
$G^{>}$.]

\begin{figure}
\includegraphics[width=\columnwidth]{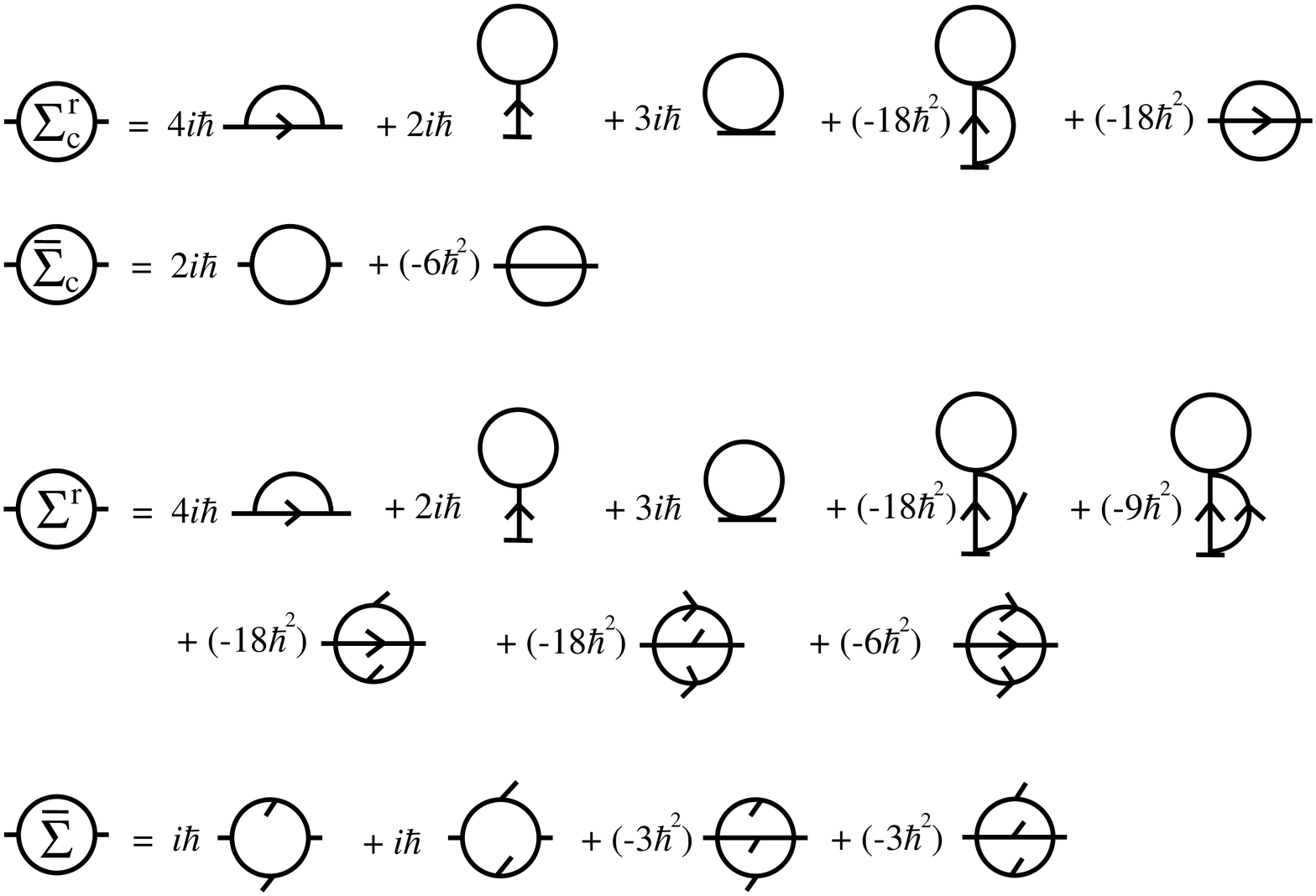}%
\caption{\label{fig-feynman-QC}The top two lines are for the
quasi-classical self-energies $\mathsf{\Sigma}_n^r$ and
$\bar{\mathsf{\Sigma}}_n$; the next two lines are the corresponding
NEGF results.  A line without arrow represents $\bar{G}$.  A line with
an arrow represents $G^r$ when read following the arrow, or $G^a$ when
read against the direction of arrow.  A line with single sided arrow
represents $G^{>}$ when following the arrow and $G^{<}$ when read
against the sense of arrow.}
\end{figure}

We compare $\mathcal{G}^r_n$ and $\bar{\mathcal{G}}_n$ with their
fully quantum-mechanical counterpart $G^r_n$ and $\bar{G}_n$ through
the nonlinear self-energies.  It is not known if a Dyson equation for
$\mathcal{G}^r_n$ is still valid in the sense that the self-energy
contains only irreducible graphs, but we simply `define' the retarded
nonlinear self-energy through
\begin{equation}
\label{eq-selfr}
\mathsf{\Sigma}_n^r = \bigl(G^{r}\bigr)^{-1} - \bigl(\mathcal{G}^{r}_n\bigr)^{-1},
\end{equation}
and similarly define $\bar{\mathsf{\Sigma}}_n$ by
\begin{equation}
\label{eq-selfbar}
\bar{\mathcal{G}}_n = \mathcal{G}^r_n \left(
\bar{\Sigma} + \bar{\mathsf{\Sigma}}_n \right) \mathcal{G}^a_n.
\end{equation}
To simplify the representation of the diagrams, we have used the fact
that $G_{jl}^r(t,t') = G_{lj}^a(t',t)$, $G_{jl}^{>}(t,t') =
G_{lj}^{<}(t',t)$, and the Keldysh relation in frequency domain
$\bar{G} = G^r \bar{\Sigma} G^a$.  In Fig.~\ref{fig-feynman-QC} we
give the lowest-order diagrams of the two types of self-energies and
contrast with the NEGF results.  Numerous cross terms involving
products of $T_{ijk}$ and $T_{ijkl}$ are not shown.  The NEGF results
are obtained in ref.~\onlinecite{WZWG-pre07} (Fig.~3) for the contour
ordered version, here we have separated out explicitly for
$\Sigma^r_n$ and $\bar{\Sigma}_n = \frac{1}{2} ( \Sigma^{>}_n +
\Sigma^{<}_n)$.  It is clear from Eq.~(\ref{eq-selfr}) and
(\ref{eq-selfbar}) that, when the nonlinear couplings $T_{ijk}$ and
$T_{ijkl}$ are zero, we have $\mathcal{G}^r_n = G^r$ and
$\bar{\mathcal{G}}_n = \bar{G}$.  Thus, for ballistic systems, NEGF
and quasi-classical MD agree exactly.  To leading order in the
non-linear couplings [$O(T_{ijk}^2)$ and $O(T_{ijkl})$] the retarded
nonlinear self-energies agree.  The difference starts only at a higher
order.  The self-energies $\bar{\Sigma}_n$ disagree even at the lowest
order.  The NEGF and quasi-classical diagrams become the same if we
take the ``classical limit'' $\hbar \to 0$ with a new definition of
classical Green's functions $G^r \to G^r_{{\rm cl}}$ and $\hbar G^{>}
\approx \hbar G^{<} \to \bar{G}_{{\rm cl}}$.  In this limit, the
distinction between $G^{>}$ and $G^{<}$ disappears.  The extra
diagrams go to zero because they are high orders in $\hbar$.

\section{Testing runs and comparison with NEGF}

\begin{figure}
\includegraphics[width=0.8\columnwidth]{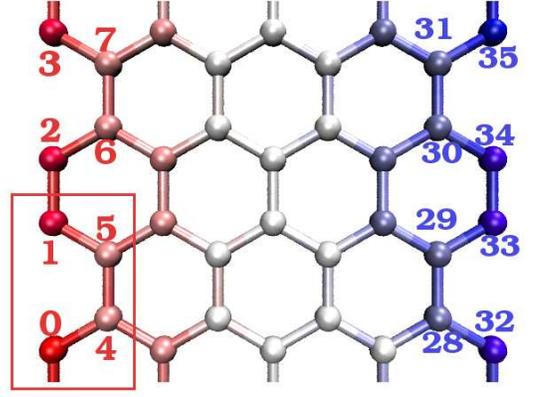}
\caption{(color online). The structure for an armchair graphene strip
with $(n, m)=(4, 2)$. The box in red line is the translational
period.}
\label{fig_cfg}
\end{figure}

Fig.~\ref{fig_cfg} is the configuration of a system in the
simulation. There are four atoms in the translational period. A pair
of numbers ($n$, $m$) are introduced to denote the number of periods
in the horizontal and vertical directions.  They should not be
confused with the chirality indices of the nanotubes.  This figure
shows the particular case of armchair graphene strip with ($n$,
$m$)=(4, 2).  For zigzag configurations, the unit cell is rotated by
90 degrees.  In the vertical direction, a periodical boundary
condition is applied.  In the simulation box, atoms in the left-most
columns labelled 0--7 are fixed left lead, atoms in the right-most
columns labelled 28--35 are the fixed right lead, and the heat baths
are applied to the columns close to them.  The temperature of the
leads are set according to $T_L = T(1+\alpha)$, and $T_R=T(1-\alpha)$.
The thermal conductance is computed from $\sigma = I_L/(T_L - T_R)$.

\begin{figure}
\includegraphics[width=\columnwidth]{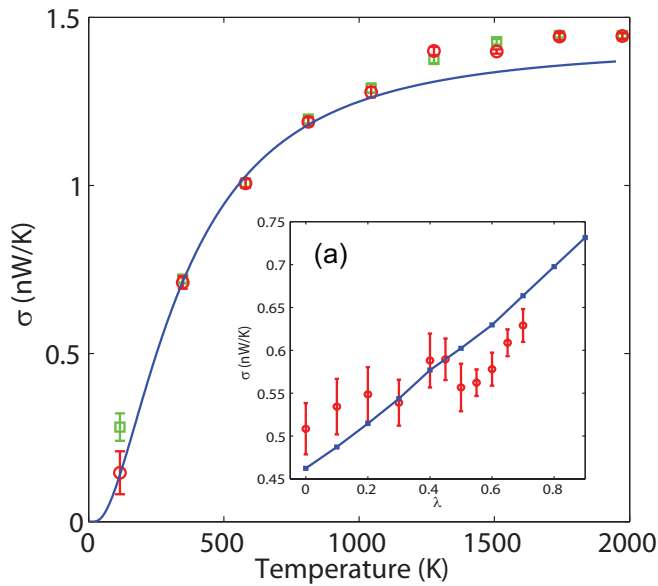}
\caption{\label{fig-ribbon}(color online). A comparison of temperature dependence of
the thermal conductance for an armchair graphene strip with $(n, m) =
(4, 2)$, solid line: NEGF, circle: QMD with velocity Verlet, square:
QMD with fourth order Runge-Kutta. (a) The $\lambda$ dependence for
the same system at 300\,K for QMD with velocity
Verlet (circle) and NEGF (solid line).}
\end{figure}

Test runs, shown in Fig.~\ref{fig-ribbon}, with parameters $\alpha =
0.4$, $\lambda = 0.6$, $f = 1.2$, $K _{\rm onsite} = 0.01$ (eV/(\AA$^2
u$)), and $\epsilon = 0.001$ ($10^{14}$ Hz) with the geometry of
Fig.~\ref{fig_cfg}, demonstrate that QMD implemented by the velocity
Verlet and fourth order Runge-Kutta gives the correct results in comparison
with ballistic NEGF.  For a system of such small sizes, the conductance behaves
ballistically.  The one implemented by the velocity Verlet agrees very
well with the NEGF result in the low-temperature regime. Other
implementation methods, like an implicit two-stage fourth order
Runge-Kutta, also turn out to give similar results.  Thus, the results
are rather insensitive to the integration algorithms used.  This
suggests the success of simulating quantum transport not only for the
one-dimensional quartic onsite model \cite{wangPRL07}, but also for
the large systems. Due to the artificial parameters added in order to
overcome the instability, the thermal conductance obtained was
slightly higher than the ballistic one in high-temperature regime.  We
note that the parameters $\lambda$, $f$, and $K_{\rm onsite}$ are
incorporated in the NEGF calculation, the effect of $\epsilon$ is not
taken care correctly in NEGF.  This may explain the discrepancy at
high temperatures.  We further analyze the $\lambda$ dependence of the
thermal conductance for the $(4, 2)$ graphene strip: the inset
Fig.~\ref{fig-ribbon}(a) represents the room temperature (300 K)
results, where the thermal conductance exhibits linear dependence on
$\lambda$.  The conductance reduces by about half when $\lambda$ is
reduced from 1 to 0.  Besides $\lambda$, other parameters also have
their own impacts, for instance, smaller $\epsilon$ lowers the effect
of the artificial damping, but requires much larger integration domain
and therefore brings the risk of truncating the spectrum and providing
the wrong self energy. The conductance is independent of $\epsilon$ if
it is in the range $0.001$ to $0.02$.

\begin{figure}
\includegraphics[width=\columnwidth]{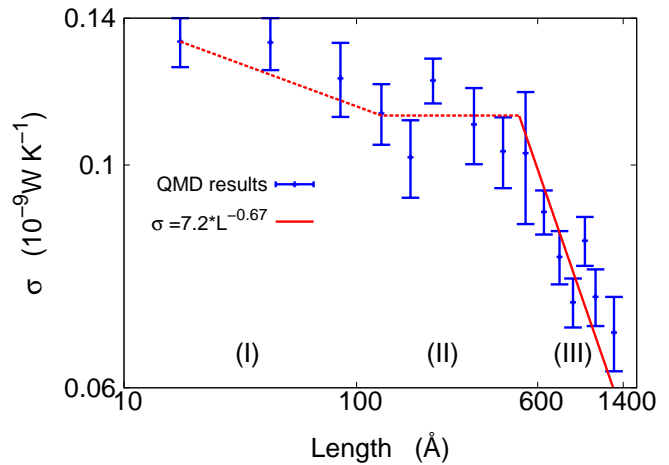}
\caption{(color online). The dependence of the thermal conductance on
  the length of the system at $300\,$K in double logarithmic
  scale. Phonon transport changes gradually from ballistic to
  diffusive with increasing length of the system.}
\label{fig_length}
\end{figure}

\begin{figure}
\includegraphics[width=\columnwidth]{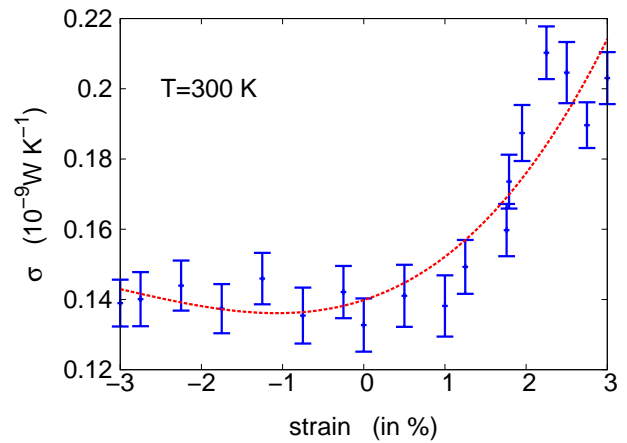}
\caption{(color online). The effect of strain on the thermal
  conductance of the graphene. The dotted line (in red) is guide to
  the eye.}
\label{fig_strain}
\end{figure}

\begin{figure}
\includegraphics[width=\columnwidth]{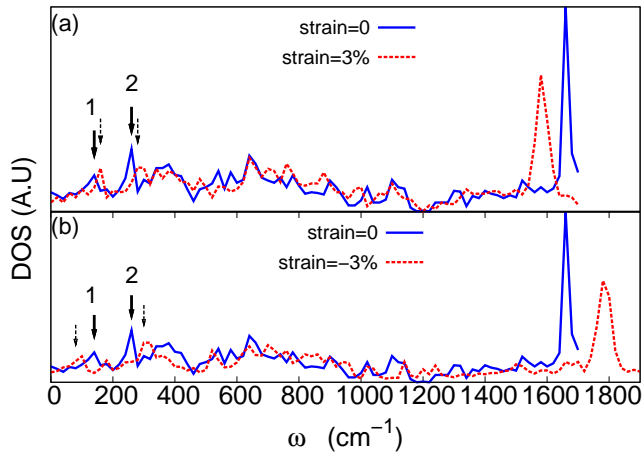}
\caption{(color online). Phonon density of state (DOS) for graphene
  under elongation strain (a), and compression strain (b).}
\label{fig_dos}
\end{figure}

\section{Results of nanoribbon under tension}

In the simulation, typical MD steps are $10^{5}$ of 0.5~fs, which is
long enough to obtain converged thermal conductance. The stabilizing
factor is $\lambda = 0.6$ with an onsite force constant for all the
atoms $K_{\rm onsite}=0.01\,$eV/(\AA$^2$u), and other parameters
$f=1.2$ and $\epsilon = 10^{-3}$ $(10^{14}$Hz).

Fig.~\ref{fig_length} shows the dependence of the thermal conductance
on the length ($L$) of the system for $(n,2)$ zigzag graphene strip at
$300\,$K. There are three regions labeled (I), (II) and (III) in the
figure respectively for $L$ in [10, 100], [100, 600] and [600,
1400]~{\AA}.  In the very short length region (I), thermal transport
should be in the ballistic regime, where thermal conductance is a
constant independent of the length of the system. But here the thermal
conductance exhibits decreasing behavior. Actually, this decreasing
behavior is mainly attributed to the localized modes at the edge of
the center region.  As shown in Fig.~\ref{fig_fixbc_edgestate}, these
modes will be more localized in longer graphene strips. So they make
little contribution to the thermal transport in short graphene strips
due to their localizing property, and even this little contribution is
further reduced with increasing length. That is the reason for the
decrease of thermal conductance in this length region. After $L\approx
100\,${\AA}, these localized modes are fully localized, so they do not
contribute to the thermal conductance anymore.  In region (II), the
thermal transport is in the ballistic regime, where the thermal
conductance is more or less a constant. In region (III), this curve
decreases as $L$ increases, indicating cross-over to the diffusive
thermal transport.  This length scale is consistent with previous
theoretical results.\cite{Mingo} Thermal conductance in this region
can be fitted by a power function $\sigma=6.1\, L^{-0.66}$ (dotted
line).  So the thermal conductivity is proportional to the length as
$L^{0.34}$. This exponent 0.34 agrees with previous results on
nanotubes and other quasi-1D systems.\cite{wang-li-PRL04,ZhangG,WangJ}

The associated values of thermal conductivity, $\kappa = \sigma L/S$,
where $L$ is length and $S$ is cross-section area, is too small.  The
smallness is attributed to the boundary resistance caused mainly by
$\lambda \neq 1$ and the omission of the delta-peak lead self
energies.  For a perfect 1D system of $(\infty, 2)$ the conductance
with the Brenner potential is 0.72 nW/K from a ballistic NEGF
calculation.  If the leads are replaced by those of omitting the
delta-peaks as discussed in Sec.III, the NEGF (4,2) system result that
is consistent with our simulation setup is reduced to $0.19\,$nW/K.
This is quite close to, but still some discrepancy with, QMD
result. These may be due to nonlinearity and other unexplained
systematic errors.

In Fig.~\ref{fig_strain}, the effect of the strain on the thermal
conductance of graphene is displayed for a system of zigzag (4, 2). To
mimic the experimental condition,\cite{Mohiuddin, Huang, Ni} the
strain is introduced in two steps. Firstly, the strain is generated to
the whole graphene system in Fig.~\ref{fig_cfg}. Secondly, atoms in
the center are fully relaxed with left and right leads fixed. And we
then do the MD simulation on this optimized graphene system. We find
that the thermal conductance increases with increasing elongation on
graphene. But compression on graphene does not change the value of the
thermal conductance appreciably.

To understand this strain effect on the thermal conductance, we study
the density of state (DOS) of the phonons in Fig.~\ref{fig_dos}. The
DOS is calculated from the Brenner empirical potential as implemented
in GULP for (4, 2) geometry with fixed boundary conditions in $x$
direction, and periodically extended in $y$ direction. We use GULP to
do optimization for the strained graphene with two leads fixed
firstly, and then calculate the DOS of this relaxed system. As shown
in Fig.~\ref{fig_dos}~(a), the high frequency Raman active mode (G
mode) around $1600\,$cm$^{-1}$ shows obvious red-shift under extension
strain, which agrees with the recent experimental
results.\cite{Mohiuddin, Huang} Furthermore, Fig.~\ref{fig_dos}~(b)
predicts the blue-shift of the G mode with compression strain.

For thermal conductance of the graphene at room temperature, the
phonon modes with frequency about $200\,$cm$^{-1}$ are important. We
can see two significant modes (1 and 2) in this frequency region in
Fig.~\ref{fig_dos}. When the graphene is elongated, both modes 1 and 2
are blue-shifted [Fig.~\ref{fig_dos}~(a)], which results in increasing
of thermal conductance. However, if the graphene is compressed, modes
1 and 2 shift in opposite directions [shown in
Fig.~\ref{fig_dos}~(b)]. As a result, the contribution of these two
modes to the thermal conductance cancels with each other. That is the
reason for the almost unchanged value of the thermal conductance under
compression.

\section{Results on nanotubes}

\begin{figure}
\includegraphics[width=\columnwidth]{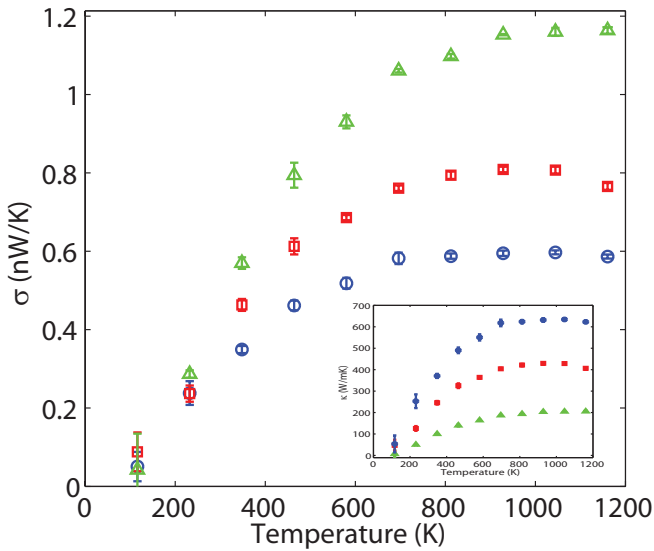}
\caption{\label{fig-zcnt}(color online). Temperature dependence of the thermal
conductance and thermal conductivity for zigzag carbon nanotubes with
$(n, m) = (10, 5)$ (triangle), $(30, 5)$ (square), and $(60, 5)$
(circle).}
\end{figure}

Fig.~\ref{fig-zcnt} shows our simulation results on zigzag carbon
nanotubes of chirality (5,0) with different lengths by using the same
parameters as that of the previous sections.  Each data point typically takes
about 48 hours on an AMD Opteron CPU.  The thermal conductivity is
computed according to $\kappa = \sigma L/S$, where $L$ is the length
of the sample, and assuming a cross-section area of $S = 12\,$\AA$^2$.
Both the thermal conductance and thermal conductivity monotonically
increase with the temperature in the low-temperature regime, which
agrees with the available experimental data and demonstrates the
ability of QMD to predict the quantum effect in this regime.  This is
completely neglected in the classical MD
approaches.\cite{berber00,maruyama02,yaozh05,ZhangG,lukes07} For
nanotubes with 12.8\,nm (30,5) and 25.6\,nm (60,5) length, as the
temperature increases, the thermal conductance and corresponding
thermal conductivity start to drop at 850K, this decrement is
consistent with the classical prediction, which indicates that the
quantum correction becomes much smaller. Yet, such decline has not been
observed in the shorter case with a length of 4.26\,nm. 
This difference shows that a transition from ballistic to diffusive
happens when the length gets longer.  The thermal conductance
decreases but conductivity still increases with nanotubes length; we
are still in transition to diffusive regime.\cite{WangJ} However,
the values at high temperatures are comparable to previous MD
results.

\section{A critique to the quantum correction method}

The quantum correction method was first suggested in
refs.~\onlinecite{czwang90,yhlee91,jli98} and used by a number of
researchers\cite{lukes07,nyang08,mchwu09,scwang09} without carefully
examining its validity.  From the simple kinetic theory of thermal
transport coefficient, the thermal conductivity can be written as
$\kappa = 1/(3V) \sum_{k} c_k v_k l_k$ where $c_k$ is the heat
capacity of a mode $k$, and $v_k$ and $l_k$ are the associated phonon
group velocity and mean free path, $V$ is the volume of the system.
Provided that the phonon velocity $v_k$ and mean free path $l_k$ of
mode $k$ are approximately independent of $k$, we can argue that the
quantum conductivity is scaled down by the quantum heat capacity from
the classical value.  In quantum correction, the temperature is also
redefined such that the classical kinetic energy is equated with the
corresponding quantum kinetic energy of a harmonic lattice.  Here it
is not clear whether the zero-point motion should be included or
not.\cite{yhlee91,lukes07,maiti97}

To a large extent, a constant phonon velocity is a good approximation.
However, it is well-known that the phonon mean free path is strongly
dependent on the frequencies, e.g., in Klemens' theory for umklapp
process\cite{klemens94}, $l \sim \omega^{-2}$.  To what extent the
quantum correction works is questionable.  There is one special case
that we can answer this question definitely, although this is for
conductance, not conductivity.  Let us consider a 1D harmonic chain
and compute its conductance exactly and compare it with a classical
dynamics with quantum correction.  The correct answer for the thermal
conductance is given by the Landauer formula,
\begin{equation}
\sigma_{{\rm QM}} = \int_{0}^{\infty} {d\omega \over 2\pi}
\hbar \omega\, T[\omega] {\partial f \over \partial T} =
\int_0^{\omega_{\rm MAX}}\!\!\!\!\! c(\omega) {d\omega \over 2\pi},
\label{sigmaQM}
\end{equation}
where the transmission $T[\omega]$ is one for a uniform chain,
$\omega_{{\rm MAX}} = \sqrt{4K/m}$ is the maximum frequency for a
chain with spring constant $K$ and mass $m$.  $c(\omega) = \hbar
\omega \partial f/\partial T$ is the heat capacity of the mode at
frequency $\omega$.  The corresponding classical value is obtained by
approximating the Bose distribution function with $f \approx k_B
T/(\hbar \omega)$. This gives the correct classical value of
conductance as
\begin{equation}
\sigma_{\rm CL} = \frac{\omega_{{\rm MAX}}}{2\pi} k_B.
\label{sigmacls}
\end{equation}
Now we consider quantum correction to Eq.~(\ref{sigmacls}).  The total
quantum heat capacity of a 1D harmonic chain is
\begin{equation}
C = \sum_k c_k = L \int_{0}^{\omega_{\rm MAX}} {c(\omega) \over v(\omega)}
{d\omega \over \pi},
\end{equation}
where $v(\omega) = d\omega/dk = (a/2) \sqrt{\omega_{\rm MAX}^2 -
\omega^2}$ is the phonon group velocity.  The classical value is $Nk_B
= Lk_B/a$, $a$ is lattice constant.

According to the quantum correction scheme, the result from a
classical dynamics is corrected by multiplying the classical value by
the ratio of quantum to classical heat capacity, given
\begin{equation}
\sigma_{\rm CORR} = \sigma_{\rm CL} \frac{C}{Nk_B} =
\int_0^{\omega_{\rm MAX}}\!
\frac{a\omega_{\rm MAX}}{\pi v(\omega)}
c(\omega) {d\omega \over 2\pi}.
\end{equation}
This does not agree with the correct quantum result of
Eq.~(\ref{sigmaQM}).  There is no need to shift the classical
temperature as $\sigma_{\rm CL}$ is independent of the temperature.
The heat capacity at frequency $\omega$ is weighted differently in two
cases.  Even if the group velocity can be approximated by a constant
by $v(0)=a\sqrt{K/m}$, valid at very low temperatures, the two results
still differ by a factor of $\pi/2$.

\section{Conclusion}

We have presented a quick derivation of the generalized Langevin
equation, emphasizing its connection with NEGF.  The inputs to run the
Langevin dynamics can be calculated in the standard way from a NEGF
phonon transport calculation.  The implementation details are given,
such as the generation of colored noise vector $\xi$.  We found quite
generically that the lead self-energies contain delta-function peaks
for quasi-one-dimensional systems.  These delta peaks represent
surface or edge modes.  This complicates the molecular dynamics
simulations.  These delta peaks in the spectra have to be removed in
order to obtain a stable simulation.  We hope that the instability is
specific to the systems of quasi-1D carbon graphene strips or
nanotubes.  If the leads are modeled as bulk 3D systems, the noise
spectra should be more smooth, and should produce a stable dynamics.
The quasi-classical approximation which results to the generalized
Langevin equation is analyzed using Feynman diagrams and its results
are compared with NEGF.  It is found that, to lead order, the nonlinear
retarded self-energy agrees with NEGF, while $\bar{\Sigma}_n$ does
not, mainly due to the fact that QMD cannot distinguish between $G^<$
and $G^>$.  As a by-product, we see easily that QMD and NEGF agree for
linear systems.  QMD also gives the correct classical limit.  We
presented test runs and compared with NEGF for the thermal
conductance.  Long $(n,2)$ graphene strips are simulated to study the
crossover from ballistic transport towards diffusive transport.
Effect of strain is also studied. The results of carbon nanotubes are
also presented.  Our simulations are one of the first examples of the
QMD on realistic systems.  Finally, the quantum correction method is
critically examined.

\section*{Acknowledgments}

The authors thank Yong Xu and Mads Brandbyge for discussion.  This
work is supported in part by research grants of National University of
Singapore R-144-000-173-101/112 and R-144-000-257-112.

\end{document}